\documentclass[12pt]{article}
\begin{document}

\title{Mixed-Mode Calculations in Nuclear Physics}
\author{V. G. Gueorguiev and J. P. Draayer \\
\\
\textit{Department of Physics and Astronomy,}\\
\textit{Louisiana State University,}\\
\textit{Baton Rouge, Louisiana 70803}}
\maketitle

\begin{abstract}
The one-dimensional harmonic oscillator in a box problem is used to
introduce the concept of a mixed-mode shell-model scheme. The method
combines low-lying ``pure mode'' states of a system to achieve
a better description in situations where complete calculations cannot be
done and the dynamics is driven by a combination of modes. The scheme
is tested for real nuclei by combining traditional spherical states,
which yield a diagonal representation of the single-particle interaction,
with collective SU(3) configurations that track deformation. An
application to the ds-shell $^{24}$Mg nucleus, using the realistic two-body
interaction of Wildenthal, is explored to test the validity of the concept.
The results shown that the mixed-mode scheme reproduces the correct binding
energy of
$^{24}$Mg (within 2\% of the exact result) as well as low-energy
configurations that have greater than 90\% overlap with exact solutions
in a space that spans less than 10\% of the full space. In the pf-shell,
the Kuo-Brown-3 interaction is used to illustrate coherent structures of
the low-lying states of
$^{48}$Cr. Alternative basis sets are suggested for future
mixed-mode shell-model studies.
\end{abstract}

Typically, two competing modes characterize the structure of a nuclear
system. One is the single-particle mode that is the underpinning of the
mean-field concept; the other is the many-particle collective behavior
manifested in the nuclear deformation. The spherical shell model is the
theory of choice when single-particle behavior dominates
\cite{Whitehead-shell model}. When deformation dominates, the Elliott SU(3)
model can be used successfully \cite{Elliott's SU(3) model}. This
manifests itself in two dominant elements in the nuclear Hamiltonian: the
single-particle term $H_{0}=\sum_{i}\varepsilon _{i}n_{i}$ and a collective
quadrupole-quadrupole interaction $H_{QQ}=Q\cdot Q$. It follows that a
simplified Hamiltonian $H=\sum_{i}\varepsilon _{i}n_{i}-\chi Q\cdot Q$ has
two solvable limits.

To probe the nature of such a system, we first consider a simpler problem:
the one-dimensional harmonic oscillator in a box of size $2L$ \cite{Armen
and Rau}. As is the case for real nuclei, this system has a finite volume
and a harmonic restoring potential, $\omega ^{2}x^{2}/2$.
Depending on the value of $E_{c}=\omega ^{2}L^{2}/2$, which plays the role
of a critical energy, there are three spectral types:

\begin{itemize}
\item[(1)]  For $\omega \rightarrow 0$ the energy spectrum is simply that of
a particle in a box.

\item[(2)]  At some value of $\omega $, the energy spectrum begins with
$E_{c}$ followed by the spectrum of a particle in a box perturbed by the
harmonic oscillator potential.

\item[(3)]  For sufficiently large $\omega $ there is a harmonic oscillator
spectrum below $E_{c}$ and a perturbed spectrum of a particle in a
box above $E_{c}$.
\end{itemize}

The last scenario (3) is the most interesting one, since it provides an
example of a two-mode system. For this case, the use of two sets of basis
vectors, one representing each of the two limits, has physical appeal for
energies around $E_{c}$. One basis set consists of the harmonic oscillator
states; the other set consists of basis states of a particle in a box. We
call this combination a mixed-mode / oblique-basis approach. In general, the
oblique-basis vectors form a nonorthogonal and sometimes an overcomplete
set. Even though a mixed spectrum is expected around $E_{c}$, our numerical
study which included up to 50 harmonic oscillator states below $E_{c}$, shows
that first-order, energy-based perturbation theory works well after a
breakdown in the harmonic oscillator spectrum.

Although the spectrum seems to be well described using first order
perturbation theory, the wave functions near $E_{c}$ have an interesting
coherent structure. For example, the zero order approximation to the wave
function used to calculate the energy may not be present in the structure of
the exact wave function. Another feature is the common shape of the
distribution that is similar to the coherent mixing observed in nuclei \cite
{VGG SU(3)andLSinPF-ShellNuclei}.

An application of the theory to the $ds$-shell nucleus $^{24}$Mg
\cite{VGG 24MgObliqueCalculations}, using the realistic two-body interaction
of Wildenthal \cite{Wildenthal}, demonstrates the validity of the mixed-mode
shell-model scheme. In this case the oblique-basis consists of the
traditional spherical states, which yield a diagonal representation of the
single-particle interaction, together with collective SU(3) configurations,
which yield a diagonal quadrupole-quadrupole interaction. The results
obtained in a space that spans less than 10\% of the full-space reproduce
the correct binding energy (within 2\% of the full-space result), as well as
the low-energy spectrum and structure of the states that have greater than
90\% overlap with the full-space results. In contrast, for a $m$-scheme
spherical shell-model calculation one needs about 60\% of the full space to
obtain comparable results.

Studies of the lower $pf$-shell nuclei $^{44-48}Ti$ and $^{48}Cr$
\cite{VGG SU(3)andLSinPF-ShellNuclei} using the realistic Kuo-Brown-3 (KB3)
interaction
\cite{KB3 interaction} show that the SU(3) symmetry breaking is due mainly
to the single-particle spin-orbit splitting. Thus the KB3 Hamiltonian could
also be considered a two-mode system. This has been further supported by the
behavior of the yrast band B(E2) values that seems to be insensitive to
fragmentation of the SU(3) symmetry. Specifically, the quadrupole
collectivity as measured by the B(E2) strengths remains high even though the
SU(3) symmetry is rather badly broken. This has been attributed to a
quasi-SU(3) symmetry \cite{Adiabatic mixing} where the observables behave
like a pure SU(3) symmetry while the true eigenvectors exhibit a strong
coherent structure with respect to each of the two bases. This provides an
opportunity for further study of the implications of two-mode calculations.

Future research may extend this to multi-mode calculations. An immediate
extension of the current scheme might use the eigenvectors of the pairing
interaction \cite{Dukelsky et al-Pairing} within the Sp(4) algebraic
approach to nuclear structure \cite{Sviratcheva-sp(4)}, with the collective
SU(3) states and spherical shell model states. Hamiltonian driven basis sets
can also be considered. In particular, the method may use eigenstates of the
very-near closed shell nuclei obtained from a full shell model calculation
to form Hamiltonian driven J-pair states for mid-shell nuclei
\cite{Heyde's-shell model}. This type of extension would mimic the
Interacting Boson Model (IBM) \cite{Iachello-1987} and the so-called
broken-pair theory. In particular, the three exact limits of the IBM \cite
{MoshinskyBookOnHO} can be considered to comprise a three-mode system.
Nonetheless, the real benefit of this approach is expected when the system
is far away of any exactly solvable limit of the Hamiltonian and the spaces
encountered are too large to allow for exact calculations.

\section*{Acknowledgments}

We acknowledge support from the U.S. National Science Foundation under Grant
No. PHY-9970769 and Cooperative Agreement No. EPS-9720652 that includes
matching from the Louisiana Board of Regents Support Fund. V. G. Gueorguiev
is grateful to the Theoretical Nuclear Physics Group of the Department of
Theoretical Physics in the Institute of Nuclear Research and Nuclear Energy
of the Bulgarian Academy of Sciences for financial help to attend the XXI
International Workshop on Nuclear Theory held June 10-15, 2002 in Rila
Mountains, Bulgaria.

\end{document}